\documentclass[epj,referee]{svjour}
\usepackage{graphics}
\onecolumn
\begin{document}
\title{Ferromagnetic resonance study of sputtered Co$|$Ni multilayers}
\author{J-M. L. Beaujour\inst{1}, W. Chen\inst{1}, K. Krycka\inst{3}, C-C. Kao\inst{3}, J. Z. Sun \inst{2} \and A. D. Kent\inst{1}}
%
\offprints{}          
\institute{Department of Physics, New York University - 4
Washington Place, New York, NY 10003, USA \and IBM T. J. Watson
Research Center - Yorktown Heights, NY 10598, USA \and Brookhaven
National Laboratory - Upton, New York 11973, USA}
\date{Received: date / Revised version: February 8$^{\mathrm{th}}$ 2007}
%
\mail {andy.kent@nyu.edu} \abstract{We report on room temperature
ferromagnetic resonance (FMR) studies of [$t$ Co$|2t$ Ni]$\times$N
sputtered films, where $0.1 \leq t \leq 0.6$ nm. Two series of
films were investigated: films with same number of Co$|$Ni bilayer
repeats (N=12), and samples in which the overall magnetic layer
thickness is kept constant at 3.6 nm (N=1.2/$t$). The FMR
measurements were conducted with a high frequency broadband
coplanar waveguide up to 50 GHz using a flip-chip method. The
resonance field and the full width at half maximum were measured
as a function of frequency for the field in-plane and field normal
to the plane, and as a function of angle to the plane for several
frequencies. For both sets of films, we find evidence for the
presence of first and second order anisotropy constants, $K_1$ and
$K_2$. The anisotropy constants are strongly dependent on the
thickness $t$, and to a lesser extent on the total thickness of
the magnetic multilayer. The Land\'e g-factor increases with
decreasing $t$ and is practically independent of the multilayer
thickness. The magnetic damping parameter $\alpha$, estimated from
the linear dependence of the linewidth, $\triangle H$, on
frequency, in the field in-plane geometry, increases with
decreasing $t$. This behaviour is attributed to an enhancement of
spin-orbit interactions with $t$ decreasing and in thinner films,
to a spin-pumping contribution to the damping.
} 
\authorrunning {Jean-Marc Beaujour}
\titlerunning {Co-Ni mutilayers}
\maketitle
\section{Introduction}
\label{intro} An understanding of magnetization dynamics in very
thin ferromagnetic layers is central to the physics and
application of spin-transfer, as devices are typically composed of
layers only a few nanometers thick \cite{spintronics}. For
example, the threshold current density for magnetic excitations is
proportional to the magnetic damping parameter $\alpha$, which can
depend on the layer environment
\cite{tserkovnyak,spintronics2,beaujour}. Further, in most
spin-transfer devices, current-induced excitation involves
precession of the magnetization out of the thin-film plane. The
easy-plane anisotropy associated with the thin film geometry
therefore plays a significant role in the resulting precession and
reversal. It is also predicted to set the threshold current, since
this anisotropy is often much larger than the in-plane anisotropy
\cite{sun}. It is therefore of interest to vary the easy-plane
anisotropy and experiment with layers with perpendicular magnetic
anisotropy \cite{sun,kent}. In fact, very recently Co$|$Ni
multilayers were incorporated into spin-torque devices
\cite{mangin}. This magnetic multilayer system is interesting
because it has a tunable easy-plane anisotropy, and devices which
incorporate such layers exhibit a reasonable GMR. Early on,
Daalderop \textit{et al.} showed that evaporated Co$|$Ni
multilayers exhibit large perpendicular magnetic anisotropy (PMA)
and perpendicular magnetization \cite{daalderop}. In addition, it
was shown that by varying the Co$|$Ni thickness ratio, it is
possible to change the effective demagnetization field. While
there has been a great deal of experimental research on Co$|$Ni
multilayers, study of the ferromagnetic resonance linewidth and
the magnetization damping of multilayers has yet to be reported.

In this paper, we present a study of the magnetic properties and
the magnetization dynamics of sputtered [$t$ Co$|2t$ Ni]$\times$N
multilayers. We discuss the sample fabrication, the structural
characterization of the films, and the experimental setup. Then
the thickness dependence of the effective anisotropy field and the
Land\'{e}e g-factor is presented and analyzed. Finally, the
magnetic damping parameter $\alpha$ is estimated and its thickness
dependence is discussed.
\section{Sample fabrication and experimental set-up}
\label{expsetup} Two series of films with the layered structure
$||$Pt$|$Cu$|$[$t$ Co$|$2$t$ Ni]$\times$N$|$Cu$|$Pt$||$ were
fabricated, where N is the number of bilayer repeats. The Co$|$Ni
layer thickness ratio was kept constant at 1 to 2, and $t$ was
varied between 0.1 and 0.6 nm. The Pt and Cu layers are 5 nm and
10 nm thick respectively. For one series of multilayers, the
number of Co$|$Ni bilayer repeats was kept constant at N$=12$ with
$0.1 \leq t \leq 0.6$ nm. Thus, the total thickness $d$ of the
magnetic multilayer varied from 7.2 to 21.6 nm. For the other set
of samples, N was chosen so that $d$ is constant at 3.6 nm with
N$=1.2/t$. Note that $d=3.6$ nm is in the range of the free layer
thicknesses used in spin-transfer devices. The multilayers were
prepared by DC magnetron sputtering at room temperature on
oxidized Si wafers, with no applied magnetic field. The base
pressure in the UHV system was $2 \times 10^{-7}$ Torr and the Ar
pressure was $1.7\times 10^{-3}$ Torr. The stacking of the
individual Co and Ni layers was achieved by opening and closing
shutters.
\subsection{Structural characterization}
\begin{figure}
\begin{center}
\resizebox{0.5\textwidth}{!}{%
  \includegraphics{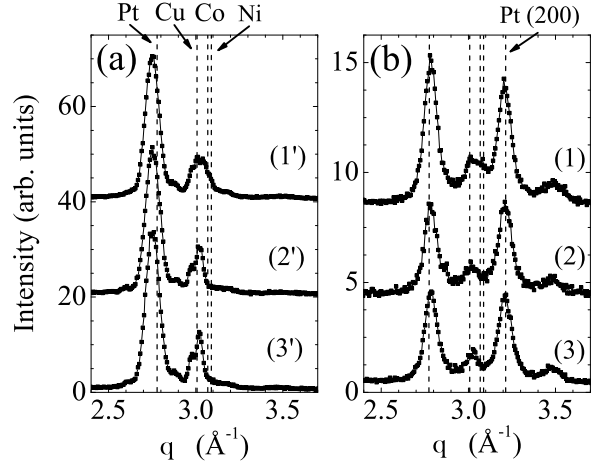}
}
 \caption{The diffraction pattern (a) at 0$^{\mathrm{o}}$ (along the [111] direction) and (b) at 54.74$^{\mathrm{o}}$
  (along the [200] direction)  for [0.1 Co$|$0.2 Ni]$\times 12$ (1, 1'), [0.3 Co$|$0.6 Ni]$\times 4$ (2, 2')
  and [0.3 Co$|$0.6 Ni]$\times 12$ (3, 3').
  The curves are shifted up for clarity. The dashed lines show the
q value for bulk fcc (111) Pt (q=2.7735 \AA$^{-1}$), Cu (q=3.0107
\AA$^{-1}$), Co (q=3.07077 \AA$^{-1}$) and Ni (q=3.0882
\AA$^{-1}$).} \label{xray01}
\end{center}
\end{figure}
Non-resonant x-ray diffraction (XRD) was carried out on films with
magnetic layer structures of [0.1 Co$|$0.2 Ni]$\times12$, [0.3
Co$|$0.6 Ni]$\times 4$, and [0.3 Co$|$0.6 Ni]$\times 12$ at 1.6295
\AA \  (7.6084 keV) at NSLS beamline X16B (Figure \ref{xray01}).
The wavelength was calibrated using an Al$_2$O$_3$ powder sample,
and the intensity was collected with a solid state Bicron
detector. In all films the Pt and Cu layers have a face centered
cubic (fcc) structure evident from the (111) and (200) reflections
present for both elements. In Figure \ref{xray01}a, the
measurements were taken with the scattering vector along the
sample normal, and here the (111) reflection dominates. In Figure
\ref{xray01}b the sample was rotated by 54.74$^{\mathrm{o}}$ about
the incident beam so that the scattering vector probed along the
[200] crystal orientation as defined for the [111] orientation
aligned with the sample normal. Although both the (111) and (200)
reflections can be seen in this geometry, there is an obvious
enhancement of the (200) reflection (second set of peaks at higher
q). These results are consistent with a textured film whose (111)
orientation aligns along the sample normal. An unexpected finding
is that while the (111) and (200) Pt peaks are close to their
nominal values at a sampling angle of 54.74$^{\mathrm{o}}$ to the
sample normal, the (111) peak shifts to lower q (expanded lattice)
when measured along the sample perpendicular. This feature is
present in all the films and indicates an expanded lattice. It is
not expected that the strain could be induced from the Cu with a
smaller lattice spacing, but the effect of in-plane strain and of
the substrate with which the Pt is also in contact with could also
play a role.

\begin{figure}
\begin{center}
\resizebox{0.5\textwidth}{!}{%
  \includegraphics{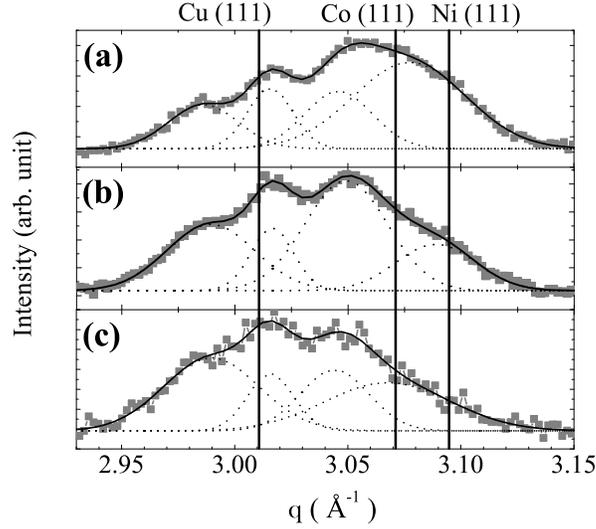}
}
 \caption{The diffraction pattern of [0.3 Co$|$0.6 Ni]$\times 12$
 (a) at the Cu K-edge (8.979 keV), (b) off resonance (9.5 keV) and (c) at the Co K-edge (7.709 keV).
 The solid black lines are the resulting Gaussian multi-peaks fit.}
\label{xray02}
\end{center}
\end{figure}

In order to quantitatively separate the Co and Ni from the Cu and
to determine whether they also have a fcc structure we obtained
resonant diffraction data from NSLS beamline X6B at both the Co
and Cu K-edges, which minimizes Co and Cu elastic scattering
contributions, respectively. For these measurements we studied the
thickest sample [0.3 Co$|$0.6 Ni]$\times$12 exclusively and
focused on the out-of-plane (111) peak region where the scattering
was strongest. Figure \ref{xray02} shows that there appears to be
four distinct peaks in this region. Two of these correspond well
to nominal Cu (111) at 3.01 \AA$^{-1}$ \ and to a mix of nominal
Co and Ni (111) (which are nearly lattice matched and are assumed
to have similar structure and texture) at 3.07 \AA$^{-1}$ to 3.09
\AA$^{-1}$. The peak between them can be explained by a
Cu$|$Co$+$Ni intermediate region centered at $3.040 \pm 0.005$
\AA$^{-1}$ \ which forms a peak distinct from the bulk Cu and bulk
Co$+$Ni. The fact that it is most prominent away from both the Cu
and Co absorption edges indicates it contains the scattering from
both elements. Finally, the lowest q peak at 2.98 \AA$^{-1}$ \
clearly decreases at the Cu K-edge (i.e. it involves Cu), does not
change relative intensity when the photon energy is tuned to the
Pt L3 edge (i.e. does not include Pt), and is unaffected by the
total thickness of the Co and Ni layers. Thus, it is likely the
result of strain on Cu by the Pt that has a slightly larger fcc
lattice structure than bulk fcc Cu.
The question remains open as to whether the intermediate
Cu$|$Co$|$Ni peak is indeed a single lattice-matched region, or
comprised of yet more overlapping sub-peaks of slightly different
lattice spacing. With the present data we cannot say with
certainty which case is true. However, the fact that the center of
this intermediate peak does not shift in reciprocal space when
going from the Co to the Cu K-edge is an indication that both the
Co and the Cu could be fully lattice matched here. A new, more
analytical method based on resonance scattering has been developed
to analyze just such situation, and will be reported in a
forthcoming article.

\subsection{Experimental technique}
\label{exptechnik} 
\begin{figure}
\begin{center}
\resizebox{0.4\textwidth}{!}{%
  \includegraphics{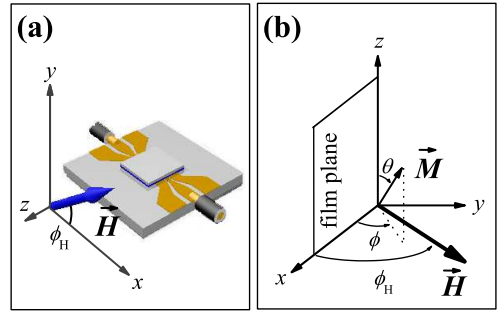}
  }
  \caption{a) Geometry of the FMR measurements.
  The Network Analyzer ports are connected directly to the CPW.
  b) Direction of the applied magnetic field $\vec{H}$ and the magnetization $\vec{M}$.
  Measurements were conducted with the field in the (\textit{x}-\textit{y}) plane, i.e $\theta = \pi / 2$.}
\label{geometry}
\end{center}
\end{figure}

FMR measurements were conducted at room
temperature employing a coplanar waveguide (CPW) as an ac magnetic
field generator and inductive sensor \cite{barry}. The CPW is made
of a 200 nm thick Au film, deposited on a semi-insulating and
polished 350 $\mu$m thick GaAs wafer. The metallic layer was
patterned using a bi-layer photolithographic process. The
microwave device was designed to have a characteristic impedance
of 50 $\Omega$ above 4 GHz: the signal line is 50 $\mu$m wide and
is separated of the ground plate by a gap of 32 $\mu$m. The CPW
was placed into a brass cavity and connected directly to the ports
of a Network Analyzer. Care was taken to avoid magnetic components
in the cavity and in all contacts to the CPW. FMR spectra were
measured by placing the magnetic sample metal face down on the
CPW. For a fixed frequency (4 to 50 GHz), the external magnetic
field was swept while measuring the S-parameters of the
transmission line. The measurements were conducted with dc fields
up to 10 kG. Figure \ref{geometry}a shows the geometry of the
measurements. The applied field was monitored with a Hall probe
sensor, and the calibration of the sensor was verified using
electron paramagnetic resonance (EPR) on
2,2-diphenyl-1-picrylhydrazyl (dpph), a spin 1/2 system. The
cavity was mounted on a rotating arm that enables FMR measurements
in the parallel geometry ($\vec{H}$ in the film plane), in the
perpendicular geometry ($\vec{H}$ normal to the film plane), and
as a function of the angle of the dc field and the film plane. All
measurements were performed with the applied field perpendicular
to the ac magnetic field.

\begin{figure}
\begin{center}
\resizebox{0.3\textwidth}{!}{%
  \includegraphics{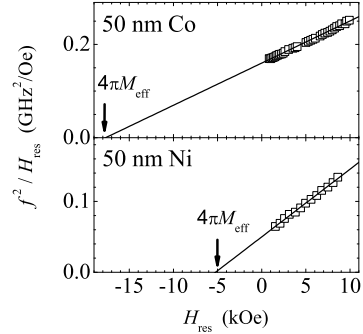}
}
  \caption{Frequency dependence of the resonance
  field $H_{\mathrm{res}}$ of
  $||$Pt$|$Cu$|$50 nm Co$|$Cu$|$Pt$||$ and $||$Pt$|$Cu$|$50 nm Ni$|$Cu$|$Pt$||$
  sputtered on Si-SiO$_2$ substrate. The frequency dependence of $H_{\mathrm{res}}$ is fitted to
  $f^2 / H_{\mathrm{res}}=(g \mu_B / \hbar) (H_{\mathrm{res}}+4 \pi M_{\mathrm{eff}})$ where the zero frequency intercept
  gives $4 \pi M_{\mathrm{eff}}$ and the slope provides the g-factor.}
\label{purefilms}
\end{center}
\end{figure}

The FMR response of a 50 nm Co and of a 50 nm Ni film sputtered on
Si-SiO$_2$ substrates was measured (Figure \ref{purefilms}). The
magnetic layers have the same layer environment as the Co$|$Ni
multilayers, with a $||$5 nm Pt$|$10 nm Cu$|$ base layer and a
$|$10 nm Cu$|$5 nm Pt$||$ top layer. The frequency dependence of
the resonance field in the parallel geometry is fitted to the
Kittel formula \cite{kittel}. For the 50 nm Co film, the estimated
value of the magnetization density
$M_{\mathrm{s}}^{\mathrm{Co}}=1408 \pm 16$ emu/cm$^3$ and the
measured g-factor $g^{\mathrm{Co}}=2.148 \pm 0.009$ agrees within
0.5\% with the parameters of the bulk material,
$M_\mathrm{s}$=1400 emu/cm$^3$ and $g=2.145$. For the film with 50
nm Ni, the experimental g-factor value $g^{\mathrm{Ni}}=2.208 \pm
0.018$ is in good agreement with the literature value ($g=2.21$),
whereas the estimated magnetization density
$M_{\mathrm{s}}^{\mathrm{Ni}}=416 \pm 14$ emu/cm$^3$ is about 15\%
smaller than that of bulk Ni. Because fcc Co and Ni have about
same lattice constant, $a_{\mathrm{Co}}=3.545$ \AA \
\cite{colatticeparam} and $a_{\mathrm{Ni}}=3.524$ \AA \
\cite{nilatticeparam} respectively and same density ($\approx$8.91
g/cm$^3$), the magnetization density of [$t$ Co$|2t$ Ni]$\times$N
is determined from the average
$M_{\mathrm{s}}=(M_{\mathrm{s}}^{\mathrm{Co}}+2M_{\mathrm{s}}^{\mathrm{Ni}})/3$,
hence $M_{\mathrm{s}}=747$ emu/cm$^3$. Note that the magnetic
materials are immiscible at room temperature \cite{su}. We assume
that the magnetization density of the individual Co and Ni layers
is thickness independent.

\section{Theory}
\label{theory} The geometry of the vectors is shown in Figure
\ref{geometry}b. $\vec M$ is the magnetization, $\vec H$ is the
applied magnetic field, and $\theta$, $\phi$ and $\phi
_{\mathrm{H}}$, are the angles associated with these vectors. The
film is in the (\textit{x}-\textit{z}) plane. We assume that the
Co and Ni layers are strongly ferromagnetically exchange coupled,
and as a consequence the magnetization of the multilayer can be
approximated as a macrospin. The applied field is chosen to remain
in the (\textit{x}-\textit{y}) plane. For a polycrystalline film
all directions in the film plane are equivalent, so $\vec{M}$ will
also remain in the (\textit{x}-\textit{y}) plane, hence
$\theta=\pi/2$. The total magnetic energy density of the system
$E_{t}$ is given by the expression \cite{chappert}:
\begin{equation}
\label{e.0} E_t = -M_{\mathrm{s}} H (\hbox{cos} \phi
\hbox{cos}\phi_{\mathrm{H}} + \hbox{sin} \phi \hbox{sin}\phi
_{\mathrm{H}}) + 2 \pi M_{\mathrm{s}}^2 \hbox{sin}^2 \phi -
(K_1+2K_2) \hbox{sin}^2 \phi + K_2 \hbox{sin}^4 \phi .
\end{equation}
The first and second term represent the Zeeman energy and the
magnetostatic energy respectively. The last two terms are the
uniaxial anisotropy energy, where $K_1$ and $K_2$ are the first
and second order effective uniaxial anisotropies. $K_1$ and $K_2$
include the surface anisotropy (N\'{e}el-type) energy and the
magneto-elastic anisotropy energy. The surface anisotropy
originates from the broken symmetry at interfaces of the
multilayer \cite{neel} and the strain can be induced by the
lattice mismatch between the layers. With our notation, positive
values of the anisotropy constants favor magnetization normal to
the film plane. For a given direction of the applied magnetic
field, the equilibrium position of the magnetization is calculated
from $(\partial E_t/ \partial \phi)=0$ and is given by the
relation:
\begin{equation}
\label{e.1} 2 H_{\mathrm{res}} \hbox{sin}(\phi _{\mathrm{H}}-\phi)
= 4 \pi M_{\mathrm{eff}} \hbox{sin} 2 \phi ,
\end{equation}
where the effective demagnetization field is defined as:
\begin{equation}
\label{e.1-2} 4 \pi M_{\mathrm{eff}} = 4 \pi M_{\mathrm{s}}
-\frac{2K_1}{M_{\mathrm{s}}}-\frac{4 K_2}{M_{\mathrm{s}}}
\hbox{cos}^2 \phi.
\end{equation}
If $K_2$ is not negligible, then $4 \pi M_{\mathrm{eff}}$ is
dependent on angle and for $K_1$ and $K_2$ positive $(4 \pi
M_{\mathrm{eff}})^{\perp}$ is larger than $(4 \pi
M_{\mathrm{eff}})^{||}$.

\subsection{Resonance field}
\label{resfield}
From the Smit and Beljers formula \cite{vonsovskii}, the resonance condition is:
\begin{equation}
\label{e.2} \omega = \gamma \sqrt{H_1 H_2},
\end{equation}
where:
\begin{equation}
\label{e.3} H_1 = H \hbox{cos}(\phi_{\mathrm{H}} - \phi) -4 \pi
M_{\mathrm{eff}}\hbox{sin}^2 \phi,
\end{equation}
and
\begin{equation}
\label{e.4} H_2= H \hbox{cos}(\phi _{\mathrm{H}} - \phi) + 4 \pi
M_{\mathrm{eff}}\hbox{cos}2 \phi + \frac{2 K_2}{M_\mathrm{s}}
\hbox{sin}^2 2\phi.
\end{equation}
$\gamma(=g \mu _B / \hbar)$ is the gyromagnetic ratio. For the the
parallel geometry ($\phi_{\mathrm{H}} = 0^{\mathrm{o}}$) and
perpendicular geometry ($\phi _{\mathrm{H}} = 90 ^{\mathrm{o}}$),
the resonance conditions are:
\begin{equation}
\label{e.5} \left(\frac{\omega}{\gamma} \right)
^2_{||}=H_{\mathrm{res}} \left(H_{\mathrm{res}}+4\pi
M_{\mathrm{s}} - \frac{2K_1}{M_{\mathrm{s}}}-\frac{4
K_2}{M_{\mathrm{s}}}\right),
\end{equation}
and
\begin{equation}
\label{e.6} \left(\frac{\omega}{\gamma}
\right)_{\perp}=H_{\mathrm{res}} - 4 \pi M_{\mathrm{s}} + \frac{2
K_1}{M_{\mathrm{s}}}.
\end{equation}

\subsection{Linewidth and damping}
\label{theorylinewidth} It is common to express the frequency
dependence of the full width at half maximum in the following form
\cite{kittel}:
\begin{equation}
\label{e.7} \triangle H (f) = \triangle H_0 + \frac{4 \pi \alpha}{\gamma
} f.
\end{equation}
$\triangle H_0$ describes an inhomogeneous broadening due to sample
imperfections and is assumed to be independent of the frequency.
The second term, known as the intrinsic linewidth, is proportional
to the magnetic damping parameter $\alpha$ and scales linearly
with the frequency $f$. By measuring the FMR signal at several
frequencies, $\alpha$ can be extracted from the slope of the curve
$\triangle H(f)$. The intercept with the zero frequency axis gives
$\triangle H_0$.

\section{Experimental data and discussion}
\label{expdata}
\subsection{Resonance field}
\label{expdatahres} 
\begin{figure}
\begin{center}
\resizebox{0.38\textwidth}{!}{%
  \includegraphics{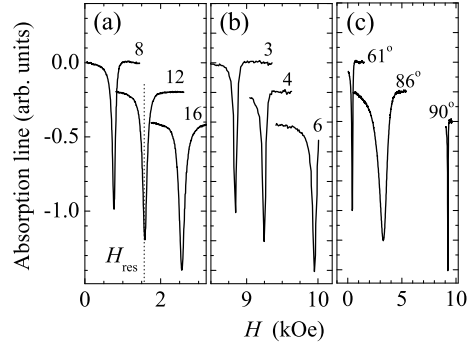}
  }
  \caption{Typical absorption lines of [0.6 Co$|$1.2 Ni]$\times 12$
  (a) at 8, 12 and 16 GHz with the applied field $\vec{H}$ in the film plane (b) at
  3, 4 and 6 GHz with $\vec{H}$ normal to the plane and
  (c) at 4 GHz for a selection of out-of-plane angles ($\phi_{\mathrm{H}}=$61$^{\mathrm{o}}$, 86$^{\mathrm{o}}$ and 90$^{\mathrm{o}}$).}
\label{lineshapes}
\end{center}
\end{figure}

Figure \ref{lineshapes} presents typical
normalized FMR peaks of a [0.2 Co$|$0.4 Ni]$\times$12 multilayer
for different field directions. The presence of a single resonance
for all the three measurements geometries suggests that the
multilayer behaves as a single magnetic film and that the
macrospin picture is appropriate to describe the FMR response. The
absorption lines were normalized by substracting the background
signal and dividing by the relative change in transmission at
resonance. The lineshape of the FMR curves is typically
Lorentzian. We also observed asymmetric lineshapes at some
frequencies, which we attribute to the mixing of the absorptive
and dispersive components of the susceptibility \cite{patton}, due
to ``poor'' deembedding of the transmission line.

\begin{figure}
\begin{center}
\resizebox{0.4\textwidth}{!}{%
  \includegraphics{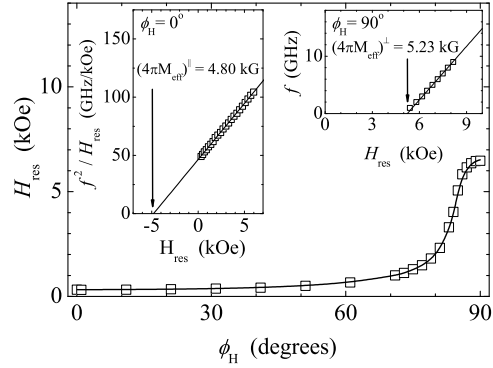}
  }
  \caption{The angular dependence of $H_{\mathrm{res}}$ of [0.3 Co$|$0.6 Ni]$\times 12$
   multilayer at 4 GHz, where $\phi_{\mathrm{H}}=0^{\mathrm{o}}$ corresponds to the parallel geometry. The
   insets show the frequency dependence of the resonance field for $\phi_{\mathrm{H}}=0^{\mathrm{o}}$ and $\phi_{\mathrm{H}}=90^{\mathrm{o}}$
   , and the corresponding effective demagnetization field $(4 \pi M_{\mathrm{eff}})^{||}$ and $(4 \pi M_{\mathrm{eff}})^{\perp}$.}
\label{typicalhres}
\end{center}
\end{figure}

The frequency dependence of $H_{{\mathrm{res}}}$ versus frequency
for in-plane and normal to the plane field directions, and the
angular dependence of the resonance field are shown in Figure
\ref{typicalhres}. The effective demagnetization in the parallel
geometry $(4 \pi M_{\mathrm{eff}})^{||} = 4801 \pm 36$ G and the
g-factor $g=2.221 \pm 0.006$ are determined using Eq. \ref{e.5}
and the method described in section 2.2. Eq. \ref{e.6} is used to
fit $H_{\mathrm{res}}$ versus $f$ in the perpendicular geometry
with $(4\pi M_{\mathrm{eff}})^{\perp}$ as the only fitting
parameter, and with the assumption that the g-factor is isotropic.
$(4\pi M_{\mathrm{eff}})^{\perp}=5230 \pm 8$ G is about 9\% larger
than $(4\pi M_{\mathrm{eff}})^{||}$. This provides evidence for a
non-negligible second order anisotropy term. Using Eq. \ref{e.5}
and \ref{e.6}, the effective uniaxial anisotropy constants
$K_1=1.551 \times 10^6$ erg/cm$^3$ and $K_2=0.080 \times 10^6$
erg/cm$^3$ are calculated. The extracted parameters are used to
fit the angular dependence of $H_{\mathrm{res}}$. As shown in
Figure \ref{typicalhres}, the fit (solid line) and the
experimental data $H_{\mathrm{res}}$ versus $\phi_{\mathrm{H}}$ at
$f=4$ GHz agree very well. In fact, we also found very good
agreement with measurements conducted at 8 and 12 GHz. When
comparing the fit and the experimental data, chi-square increases
by a factor 2 when $K_2$ is set to zero. $K_1$ and $K_2$ are
positive and $K_1
>> K_2$ which means that they favor the magnetization out of the
film plane. Nevertheless, it is clear from the shape of the curve
$H_{\mathrm{res}}(\phi_{\mathrm{H}})$, where
$(H_{\mathrm{res}})^{\perp}
> (H_{\mathrm{res}})^{||}$, that the preferential direction for
the magnetization of the multilayer is in the film plane.

\begin{figure}
\begin{center}
\resizebox{0.35\textwidth}{!}{%
  \includegraphics{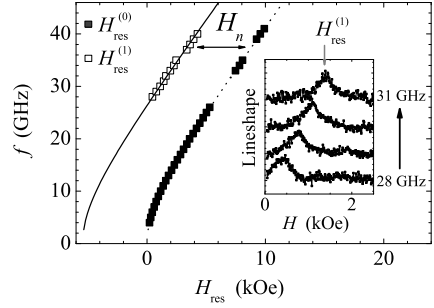}
  }
  \caption{The frequency dependence of the resonance field of
  the uniform mode $H_{\mathrm{res}}^{(0)}$ (filled symbols) and of the
  spin wave mode $H_{\mathrm{res}}^{(1)}$ (open symbols) of the multilayer
  [0.6 nm Co$|$1.2 nm Ni]$\times$12 in the parallel geometry. The peak separation is $H_n \simeq 5.4$ kOe. The insert shows the absorption line
  of the high frequency resonance peak at $f=$28, 29, 30 and 31 GHz.}
\label{psw}
\end{center}
\end{figure}

For frequencies above 25 GHz, an additional resonance peak is
observed in the absorption line of the thickest magnetic
multilayer [0.6 Co$|$1.2 Ni]$\times 12$ (inset of Figure
\ref{psw}). The frequency dependence of this resonance peak
suggests that it is associated with a spin-wave resonance mode.
The high-order resonance field $H_{\mathrm{res}}^{(1)}$ and its
amplitude are smaller than that of the uniform resonance mode. In
a model proposed by Kittel \cite{kittel2}, the field splitting
with respect to the uniform mode is described by $H_n=2 A k^2
/M_{\mathrm{s}}$, where $A$ is the exchange constant. If there is
no pinning of the surface spins at the bottom and top interfaces
then $k=(n-1) \pi/d$, where $n$ is the perpendicular standing spin
wave number and $d$ is the total thickness of the magnetic film.
Using a weighted value for the exchange constant of the Co$|$Ni
multilayer $A=(A^{\mathrm{Co}}+2A^{\mathrm{Ni}})/3$ with
$A^{\mathrm{Co}}=1.3 \times 10^{-6}$ erg/cm \cite{tannenwald} and
$A^{\mathrm{Ni}}=0.75 \times10^{-6}$ erg/cm \cite{martin}, we find
$n=1.96 \pm 0.03 \approx 2$, which corresponds to the surface spin
wave of the 1$^{\mathrm{st}}$ order. The Kittel model assumes that
the magnetization density is uniform across the film thickness.
Such a spin wave mode is therefore not expected to be observed
with a uniform ac field since $\sum \vec{m_z}=0$. Using
approximations for the conductivity of the magnetic multilayer and
its permeability at high field (off resonance), the ac field
attenuation through the multilayer [0.6 nm Co$|$1.2 nm Ni]$\times
12$ is estimated to be about 3\% at 30 GHz. We suspect that
inhomogeneity of the ac field and the structural asymmetry of our
Co$|$Ni multilayer film, with two different outer interfaces,
Cu$|$Co and Ni$|$Cu, can lead to $\sum m_z \neq 0$, and excitation
of spin-wave modes.

\begin{figure}
\begin{center}
\resizebox{0.4\textwidth}{!}{%
  \includegraphics{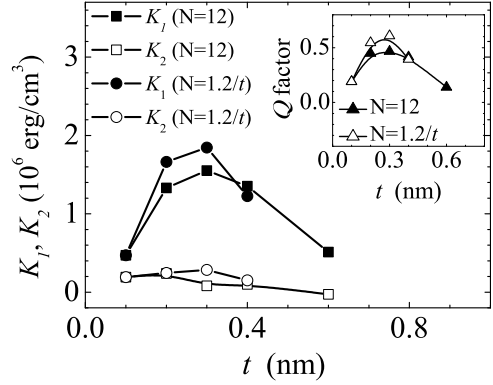}
  }
  \caption{Thickness dependence of the anisotropy constants $K_1$ and $K_2$ and of the quality factor $Q$. The solid lines are guides-to-the-eye.}
\label{tdependenceQK}
\end{center}
\end{figure}

The thickness dependence of the effective anisotropy constants of
the multilayers [$t$ Co$|2t$ Ni] with 12 repeats and with 1.2/$t$
repeats is shown in Figure \ref{tdependenceQK}. In the thickness
range $0.1 \leq t \leq 0.6$ nm, $K_1$ is positive and greater than
$K_2$, and it exhibits a maximum value of about 1.8$\times 10^6$
erg/cm$^3$. $K_1$ of the multilayers with 12 repeats increases by
a factor 3 when $t$ decreases from 0.6 nm to 0.3 nm. The second
order anisotropy term is negative for $t=0.6$ nm ($K_2=-0.028
\times 10^6$ erg/cm$^3$) and changes to positive values for
smaller $t$. For $0.1 \leq t \leq 0.4$ nm, $K_1$ and $K_2$ of
multilayers of 12 and 1.2/$t$ \ Co$|$Ni repeats have very similar
$t$ dependence.

\begin{figure}
\begin{center}
\resizebox{0.35\textwidth}{!}{%
  \includegraphics{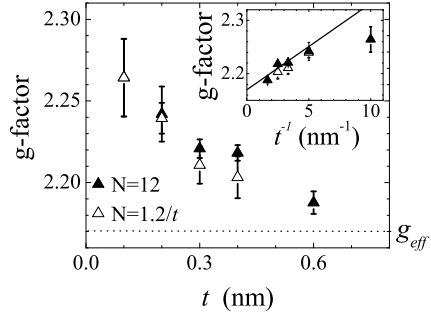}
  }
  \caption{Thickness dependence of the g factor for Co$|$Ni multilayers of 12 bilayer repeats (filled symbols),
  and multilayers with constant layer thickness (open symbols). The inset shows $g$ as a function of $1/t$.
  The solid line is the best fit of $g(1/t)$ for films with N=12 and without taking into account the data point at $1/t=10$ nm$^{-1}$.}
\label{tdependenceg}
\end{center}
\end{figure}

In Figure \ref{tdependenceg} the g-factor as a function of the
thickness $t$ is presented. The Land\'{e} g-factor of the films
with same magnetic layer thickness (3.6 nm) varies with $t$ in the
same way as that of the multilayers with constant number of
repeats. This suggests that $g$ does not depend on the total
thickness of the multilayer, and it is the thickness of the Co and
Ni layers that induces the change in $g$. The dashed line in
Figure \ref{tdependenceg} shows the calculated value of the
g-factor of the multilayers based on a geometric average
\cite{wangsness}:
\begin{equation}
\label{e.15} g_{\mathrm{eff}} =
\frac{M_{\mathrm{s}}^{\mathrm{Co}}+2M_{\mathrm{s}}^{\mathrm{Ni}}}{M_{\mathrm{s}}^{\mathrm{Co}}/g^{\mathrm{Co}}
+2M_{\mathrm{s}}^{\mathrm{Ni}}/g^{\mathrm{Ni}}}.
\end{equation}
Using the values found for the 50 nm Co and 50 nm Ni films, we
find $g_{\mathrm{eff}}=2.170 \pm 0.058$.
The inset Figure \ref{tdependenceg} shows $g$ versus $1/t$. The
data points fall on a line for $t \geq 0.2$ nm. The intercept of
the linear fit with the $t^{-1}$=0 axis gives $g=2.170 \pm 0.012$
which agrees with the value of $g_{\mathrm{eff}}$. A similar
linear fit for the film with N=($1.2/t$) gives $g=2.163 \pm
0.024$.

\subsection{Discussion of resonance field}
\label{hresdiscussion} In the study of magnetic films with large
PMA, the dimensionless quality factor $Q$ is often used as a
figure of merit. $Q$ is defined by the ratio of the uniaxial
anisotropy to the demagnetizing energy. In the limit that the PMA
field overcomes the demagnetization field, the normal to the film
plane becomes the axis of easy magnetization and $Q> 1$. The inset
of Figure \ref{tdependenceQK} shows the dependence of
$Q=(K_1+K_2)/2 \pi M_{\mathrm{s}}^2$ as a function of $t$. The
quality factor of the Co$|$Ni sputtered multilayers does not
exceed 0.7. The magnetization of the films remains preferentially
in the film plane, in contrast to what has been observed for
evaporated multilayers of the same Co:Ni thickness ratio
\cite{bloemen}. The result can be in parts explained by the
crystallographic growth direction of the underlayer that does not
promote a highly textured (111) film. The diffraction pattern of
the multilayers exhibit a (111) Pt peak, but also a peak that
corresponds to Pt (200). Zhang $et$ $al.$ conducted XRD and
magnetometry measurements on sputtered Co$|$Ni grown on Au and Ag
base layer \cite{zhang}. The authors found that only the
multilayers grown on Au have the magnetization easy axis normal to
the film plane. The XRD study showed that the Ag (111) intensity
peak is 10 times smaller than that of Au (111), with the presence
of a second peak of intensity corresponding to Ag (200).

The magneto-elastic contribution arises from the strain induced by
the lattice mismatch $\eta$ between adjacent layers. The largest
strain originates from the lattice mismatch at the Cu$|$Co
interface where $|\eta _{\mathrm{Cu}|\mathrm{Co}}| \approx 1.8$\%,
compared to $|\eta _{\mathrm{Co}|\mathrm{Ni}}| \approx$0.6\% at
the Co$|$Ni interface. The anisotropy constants of the films with
same multilayer thickness depends strongly on $t$. Therefore, the
interfaces Cu$|$Co and Ni$|$Cu do not appear to play a significant
role in the $t$ dependence of the anisotropy constants.
Consequently, the variation of the anisotropies with $t$ is
associated with the Co$|$Ni interfaces, as predicted by Daalderop
\textit{et al.} \cite{daalderop}. From \textit{ab-initio}
calculations, the authors found that in the multilayers, the
interface anisotropy is controlled by d-state occupancy near the
Fermi level at the Co$|$Ni interface. The presence of an optimum
thickness $t$ for large $K_1$ value can be understood as follows.
With increasing $t$, the number of Co$|$Ni interfaces per unit
volume decreases and $K_1$ decreases. In the limit of very small
$t$, the magnetic layers are not uniform. The multilayer breaks
down into an alloy-type structure and $K_1$ decreases.

The g-factor is related to the ratio of the orbital to spin
moments $\mu_{\ell}/ \mu_s$ by $\mu_{\ell}/ \mu_s=(g-2)/2$
\cite{farle}, hence one deduces that $\mu _{\ell} / \mu_s$
increases by 3.4\% when $t$ is decreased from 0.6 nm to 0.1 nm.
This enhancement depends on the thickness of the individual Co and
Ni layers. Significant increase of the ratio $\mu_\ell / \mu_s$,
up to 40\% have been reported for ultrathin FM in contact with a
NM \cite{mulmus}. The enhancement was attributed to the breaking
of the symmetry at the interface, where the orbital moment of the
surface/interface atoms is enhanced compared to that of the bulk
atoms.

\subsection{Linewidth and magnetic damping}
\label{expdatalinewidth} 
\begin{figure}
\begin{center}
\resizebox{0.35\textwidth}{!}{%
  \includegraphics{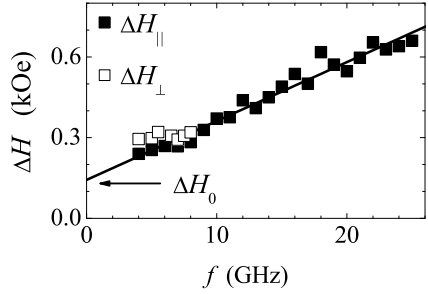}
  }
  \caption{Frequency dependence of the linewidth in the parallel ($\triangle H_{||}$)
  and perpendicular ($\triangle H_{\perp}$) geometry of the film [0.2 nm Co$|$0.4 nm Ni]$\times 6$.
   The extrinsic contribution $\triangle H_0$ and the intrinsic contribution $\triangle H/df$ to
   the linewidth are extracted from the linear best fit of $\triangle H_{||}(f)$ (solid line).}
\label{linewidthtypical}
\end{center}
\end{figure}

\begin{figure}
\begin{center}
\resizebox{0.34\textwidth}{!}{%
  \includegraphics{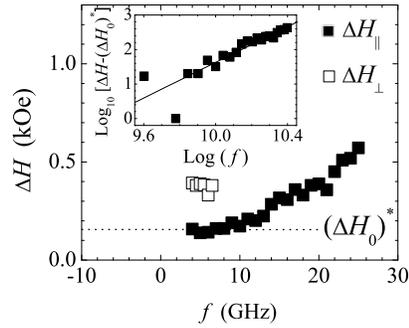}
  }
  \caption{The linewidh versus frequency for the film
  [0.1 Co$|$0.2 Ni]$\times 12$ in the parallel geometry.
  $(\triangle H_0)^{*}$ is the extrapolated linewidth when $f$ approaches zero
  frequency linewidth. The inset shows the data in
Log$_{10}$-Log$_{10}$ scale, and the linear best fit (solid
line).} \label{linewidthpeculiar}
\end{center}
\end{figure}

\begin{figure}
\begin{center}
\resizebox{0.35\textwidth}{!}{%
  \includegraphics{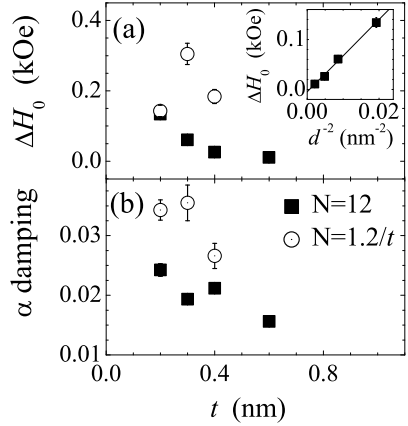}
    }
  \caption{Dependence of the extrinsic contributions $\triangle H_0$ and the damping $\alpha$
  extracted from the slope $d \triangle H /df$ for multilayers with same number
  of bilayer repeats (N=12) and for films with same thickness (N=1.2/$t$). The inset
  in (a) shows $\triangle H_0$ versus 1/$d^2$ for the multilayers with N=12, where $d$
  is the total multilayer thickness. The solid line is the linear best fit of $\triangle H_0$ versus 1/$d^2$.}
\label{dampingdeltah0}
\end{center}
\end{figure}

The linewidth was studied as a function
of frequency in the parallel and the perpendicular geometry.
Figure \ref{linewidthtypical} shows typical data for [0.2 nm
Co$|$0.4 nm Ni]$\times 12$ multilayer.
The linewidth in the parallel geometry $\triangle H_{||}$ and in
perpendicular geometry $\triangle H_{\perp}$ follows similar
frequency dependence, and increases with frequency. $\triangle
H_{\perp}$ is slightly larger than $\triangle H_{||}$ at low
frequencies (4-6 GHz), which we attribute to a small misalignment
of the applied field with the normal to the film plane. Exchange
narrowing can also lead to $\triangle H_{\perp} > \triangle
H_{||}$ \cite{hurdequint}. $\triangle H$ in the parallel geometry
depends linearly on frequency, with a zero frequency offset. The
inhomogeneous broadening $\triangle H_0=143 \pm 16$ Oe and the
damping parameter $\alpha=0.0343 \pm 0.0017$ are extracted from
the best linear fit (Eq. \ref{e.7}). Note that the inhomogeneous
ac field due to the finite width of the transmission line can lead
to the broadening of the linewidth \cite{counil}. The additional
linewidth increases with the magnetic film thickness and is
inversely proportional to the frequency. For the thickest magnetic
multilayer ($d=21.6$ nm), we estimate the additional linewidth at
5 GHz to be less than 1$\%$ of the measured linewidth. Spin wave
contributions to the linewidth from the CPW geometry can therefore
be neglected. The dependence of $\triangle H_{||}$ versus
frequency is linear for most samples. Only the thinnest film [0.1
nm Co$|$0.2 nm Ni]$\times 12$ exhibits a non linear dependence on
frequency so that $\alpha$ and $\triangle H_0$ could not be
determined (Figure \ref{linewidthpeculiar}). A linear fit of
$\triangle H_{||}$ as a function of frequency in a
Log$_{10}$-Log$_{10}$ plot scale gives $n=2.58 \pm 0.30$ with
$\triangle H_{||} \propto f^n$. The behaviour might originate from
two-magnon scattering contribution to the linewidth induced by the
film roughness. Indeed, Twisselmann \textit{et al.} found that a
64 nm Py film grown on a highly oriented roughness shows a non
linear in-plane linewidth frequency dependence, when the dc field
is applied perpendicular to the scratches \cite{twisselmann}.
The magnetic damping parameter and the extracted inhomogeneous
contribution $\triangle H_0$ to the linewidth of the multilayers
is shown in Figure \ref{dampingdeltah0}. $\alpha$ and $\triangle
H_0$ of the films with N=12 increases monotonically with
decreasing $t$. The thickness dependence $t$ of $\triangle H_0$ of
the films with $N=1.2/t$ is not clear. The inset of Figure
\ref{dampingdeltah0}a shows $\triangle H_0$ as a function of
$1/d^{2}$, where $d$ is the total thickness of the magnetic
multilayer. The extrinsic contribution to the linewidth increases
linearly with $d^{-2}$. The damping of the films with $t=0.2$ and
0.3 nm and thickness 3.6 nm is 40\% larger than that of the film
with $t=0.4$ nm. The largest damping values are found for the
films 3.6 nm thick, which are the thinnest magnetic multilayers.

\subsection{Discussion of Enhanced Damping}
\label{discussionlinewidth} 
\begin{figure}
\begin{center}
\resizebox{0.4\textwidth}{!}{%
  \includegraphics{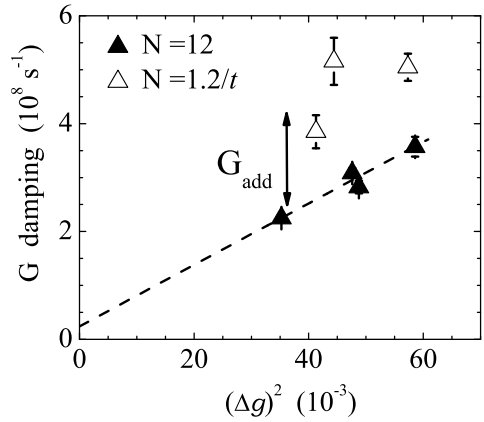}
}
  \caption{The Gilbert damping constant G as a function of $(\triangle g)^2=(g-2.0023)^2$.
   $\mathrm{G}_{\mathrm{add}}$ is the additional damping from spin
   pumping: the difference between the damping of the thickest film ([0.6 nm Co$|$1.2 nm Ni]$\times 12$)
   and that of the films of 3.6 nm thickness. The dashed line is a guide to the eye.} \label{Gversusg}
\end{center}
\end{figure}

The enhancement of the magnetic damping
constant in thin magnetic films can originate from several
mechanisms. It is generally believed that the spin-orbit
interaction in a ferromagnet, which couples the spin to the
lattice, plays a dominant role in the damping mechanism. The
following expression was theoretically derived \cite{elliott}:
\begin{equation}
\label{e.13} \hbox{G} \propto \triangle g^2,
\end{equation}
where $\triangle g$ is the deviation of the Land\'{e} g-factor
from the free electron value 2.0023 and G is the Gilbert damping
constant defined by G$=\alpha \gamma M_{\mathrm{s}}$. Figure
\ref{Gversusg} shows the Gilbert damping constant versus
$(\triangle g)^2$ for the two series of multilayers. For a given
Land\'{e} g-factor value, the damping is largest for the film 3.6
nm thick.

A mechanism that can explain the enhanced Gilbert damping in the
3.6 nm thick multilayer films is the spin-pumping
\cite{tserkovnyak2}. The precessing magnetization of the Co$|$Ni
multilayers generates a spin current that flows through the
adjacent Cu layers and relaxes in Pt, a strong spin scatterer. As
a consequence, the damping is enhanced. For a symmetric structure
where the FM is embedded between $||$NM2$|$NM1$|$ bilayers, the
enhanced damping $\alpha_{\mathrm{eff}}$ is:
\begin{equation}
\label{e.16} \alpha_{\mathrm{eff}} = \alpha _0 + \frac{g \mu _B}{2
\pi M_{\mathrm{s}}} \frac{\tilde{g}^{\uparrow
\downarrow}_{\mathrm{eff}} S^{-1}}{d}.
\end{equation}
$\alpha _{0}$ is the residual damping (bulk) and the second term
represents the additional damping from spin pumping where
$\tilde{g}_{\mathrm{eff}} ^{\uparrow \downarrow} S^{-1}$ is the
effective spin mixing conductance. Eq. \ref{e.16} is valid when
the NM2 is a perfect spin sink and the thickness of the NM1 is
much smaller than the spin diffusion length of the material. Pt is
known to act as a perfect spin sink and the Cu layer of 10 nm is
much thinner than the spin diffusion length at room temperature
$\lambda _{\mathrm{sf}}=350$ nm \cite{jedema}. The additional
damping induced by spin pumping effect in the 3.6 nm film is the
difference between G of the thickest multilayer film (21.6 nm) and
that of the 3.6 nm films (Figure \ref{Gversusg}). Using an average
value for the g-factor, we obtain $\alpha _{\mathrm{add}} \approx
0.015$. The effective spin mixing conductance of sputtered Co
layer embedded between $||$Pt$|$Cu$|$ bilayer was found to be
$g_{\mathrm{eff}}^{\uparrow \downarrow} S^{-1}=(1.63 \pm 0.18)
\times 10^{15}$ cm$^{-2}$ \cite{beaujour2}. The additional damping
computed using Eq. \ref{e.16} is $\alpha_{\mathrm{add}} \approx
0.020$. This is in the range of the value found using the
experimental data.

\section{Conclusion}
This FMR study on several sputtered $t$ Co$|2t$ Ni multilayers
shows in-plane preferential orientations, and also perpendicular
effective anisotropy $K_1$ as large as $2.8 \times 10^6$
erg/cm$^3$. We have provided evidence that to understand the
frequency dependence and angular dependence of the resonance
field, second order anisotropy terms have to be considered. The
Land\'{e} factor increases with decreasing $t$ and the enhancement
depends on the thickness of the individual Co and Ni layers. The
thickness dependence of $g$ is explained in terms of the lowering
of the symmetry at the Co$|$Ni interface. The extrinsic and the
intrinsic contribution to the FMR linewidth increases with
decreasing thickness $t$ of the individual layers. $\triangle H_0$
follows a 1/$d^2$ dependence. The enhancement of the magnetic
damping is attributed to the increase of spin-orbit interaction
and to the spin-pumping. In order to clarify this point it would
be interesting to study similar magnetic multilayers without the
Pt layers. Indeed, without adjacent layers with strong spin-orbit
scattering, the additional damping is expected to be weak.

The possibility to tune the easy plane anisotropy by changing the
thickness of the individual magnetic layers makes the Co$|$Ni
multilayer an interesting magnetic structure to be integrated in
spin-transfer devices. In addition, the damping while larger than
that of Permalloy is similar to that of Co ultrathin films. It
would also be interesting to compare the high frequency dynamics
of films grown with different underlayers and deposition methods.

\label{conclusion}
\begin{acknowledgement}
We thank Dr. G. de Loubens for fruitful discussions. This research
is supported by NSF-DMR-0405620. Use of the National Synchrotron
Light Source, Brookhaven National Laboratory, was supported by the
U.S. Department of Energy, Office of Science, Office of Basic
Energy Sciences, under Contract No. DE-AC02-98CH10886.
\end{acknowledgement}

\end{document}